\documentclass[11pt]{article}
\usepackage{moriond,epsfig}

\def\be{\begin{equation}}
\def\ee{\end{equation}}
\def\bea{\begin{eqnarray}}
\def\eea{\end{eqnarray}}

\begin{document}
\vspace*{4cm}
\title{New Strange Asymmetry Results from NuTeV}

\author{ D. ~Mason \address{University of Oregon, Eugene OR 97403} for the NuTeV Collaboration}

\maketitle\abstracts{
Results from the next to leading order (NLO) dimuon analysis  
from the NuTeV experiment at FNAL are
presented. Charged current interactions in neutrino-nucleon scattering
with two oppositely charged muons in the final state allow
direct study of charm production and measurement of the strange sea.
NuTeV's sign selected beam gives it the ability to extract the strange and
antistrange seas independently, for which an asymmetry has been predicted
in some theoretical models, and which is currently of intense interest in
interpreting neutrino electroweak results.  The dimuon results presented
here represent the first analysis of dimuon events performed utilizing
DISCO, a fully NLO cross section code differential in all variables required
to model detector acceptance.
}

\section{Introduction}

Events containing two oppositely signed muons (dimuons) from muon neutrino deep inelastic scattering experiments provide a unique window into the strange quark content of the nucleon\cite{NREV}.  These events occur in charmed particle production from a charged current (CC) interaction with a strange (or down) quark.  Approximately 10\% of the time the charmed hadrons decay semi-muonically, yielding a final state with two oppositely charged muons, one from the weak vertex, one from the charm decay.  Charm production from down quarks is Cabibbo suppressed, making dimuons most sensitive to the strange sea.  Dimuons are clearly distinguishable in a large neutrino detector such as that employed by NuTeV \cite{NIM}.

NuTeV ran during FNAL's 96-97 fixed target run and recorded
5102 dimuon events from CC $\nu_\mu$, and 1458 from CC $\overline{\nu}_\mu$ scattering in its iron target.   The detector was calibrated throughout the run by muon, electron, and hadron beams so its response is well understood.   NuTeV's beamline was constructed to be able to select $\nu_\mu$ or $\overline{\nu}_\mu$ beams with high purity.  This a priori knowledge of whether events were the result of a neutrino or antineutrino interaction, unique to NuTeV, allows one to measure the strange sea independently from the antistrange sea.  

Of particular topical interest is whether the strange sea is different from the antistrange sea.  It has been proposed, for example, that the proton wavefunction could have a virtual $K^+ \Lambda$ pair component which would lead to an asymmetry between strange and antistrange seas\cite{brodsky}.  More recently this possibility has been entertained\cite{Gambino,Rapid,Kretzer:2003wy}
as an explanation for the almost 3$\sigma$ difference between NuTeV's $\sin^2\theta_W$ result\cite{Sam} and global fits.  To eliminate this discrepancy a strange-antistrange asymmetry, as defined by the momentum weighted integral: $S^-\equiv\int x[s(x)-\overline{s}(x)] dx$, would need to be positive and as large as +0.007\cite{McFarland:2003jw}.  NuTeV's dimuon data sample offers the opportunity to address this question directly.


\section{Fitting the Dimuon Cross Section}
NuTeV has made its dimuon data available in the form of a forward dimuon cross section table\cite{Max}.  The cross section of dimuons with charm decay muons of energy greater than 5 GeV was extracted in bins of Bjorken $x$, inelasticity $y$, and neutrino energy $E_\nu$, and corrected for detector smearing effects.  

A fit to the dimuon cross section table is made up of the following components:
\begin{equation}
\frac{d\sigma_{charm}}{dxdy} ~ \cdot~ EMC \cdot~ B_c ~\cdot ~\mathcal{A}~ =\!\!\fbox{fit}\!\!\Rightarrow~ \frac{d\sigma_{2\mu}}{dxdy}
\end{equation}
Where $\frac{d\sigma_{2\mu}}{dxdy}$ is the measured dimuon cross section table mentioned above, to which the left side of the expression above is compared.   $\frac{d\sigma_{charm}}{dxdy}$ is the cross section model, dependent on the strange sea and charm mass, which would be varied in a fit.  $EMC$ is an $x$ dependent correction for the EMC effect and nuclear shadowing, parameterized from a fit to data from charged lepton scattering experiments in heavy nuclear targets and deuterium\cite{SELIG}.  $B_c$ is the semileptonic charm branching ratio ($0.093\pm0.008$, from a re-analysis of FNAL E-531\cite{TAB}), and $\mathcal{A}$ is a kinematic acceptance correction accounting for the 5 GeV cut on the energy of the muon from the semileptonic charm decay. $\mathcal{A}$ depends on $E_\nu$, $x$, $y$ as well as fragmentation and, at NLO, charm mass.  A Monte Carlo simulation using the DISCO\cite{KMO} NLO cross section code, differential in all variables required to model this acceptance, was used to calculate $\mathcal{A}$.  



Fits were performed using acceptance corrections calculated with a constant Collins-Spiller\cite{CSPILLER} epsilon of 0.75, since the charm mass and epsilon are correlated at NLO.  It was determined that the strange asymmetry was insensitive to the choice of epsilon, and a value where NLO Monte Carlo matched to data well was chosen, and varied over a conservative range (0.25) for error estimates.  The charm mass is then allowed to vary in the fit.

\section{Strange Asymmetry Results}

Several extractions of the strange and antistrange seas from the dimuon cross section tables have been performed at different orders in QCD (LO and NLO) as well as with different methods of parameterization.  The first class of fits treat the strange and antistrange seas completely independently, and use a more or less traditional parameterization:

\vspace{-0.1in}
\begin{eqnarray}
  s(x,Q) & = & \kappa (1-x)^{\alpha} \left[\frac{\overline{u}(x,Q) + \overline{d}(x,Q)}{2}\right] \\
  \overline{s}(x,Q) & = & \overline{\kappa} (1-x)^{\overline{\alpha}} \left[\frac{\overline{u}(x,Q) + \overline{d}(x,Q)}{2}\right]
\end{eqnarray}

\begin{figure}[h!]
\begin{center}
\resizebox{3.0in}{2.3in}
   {\includegraphics{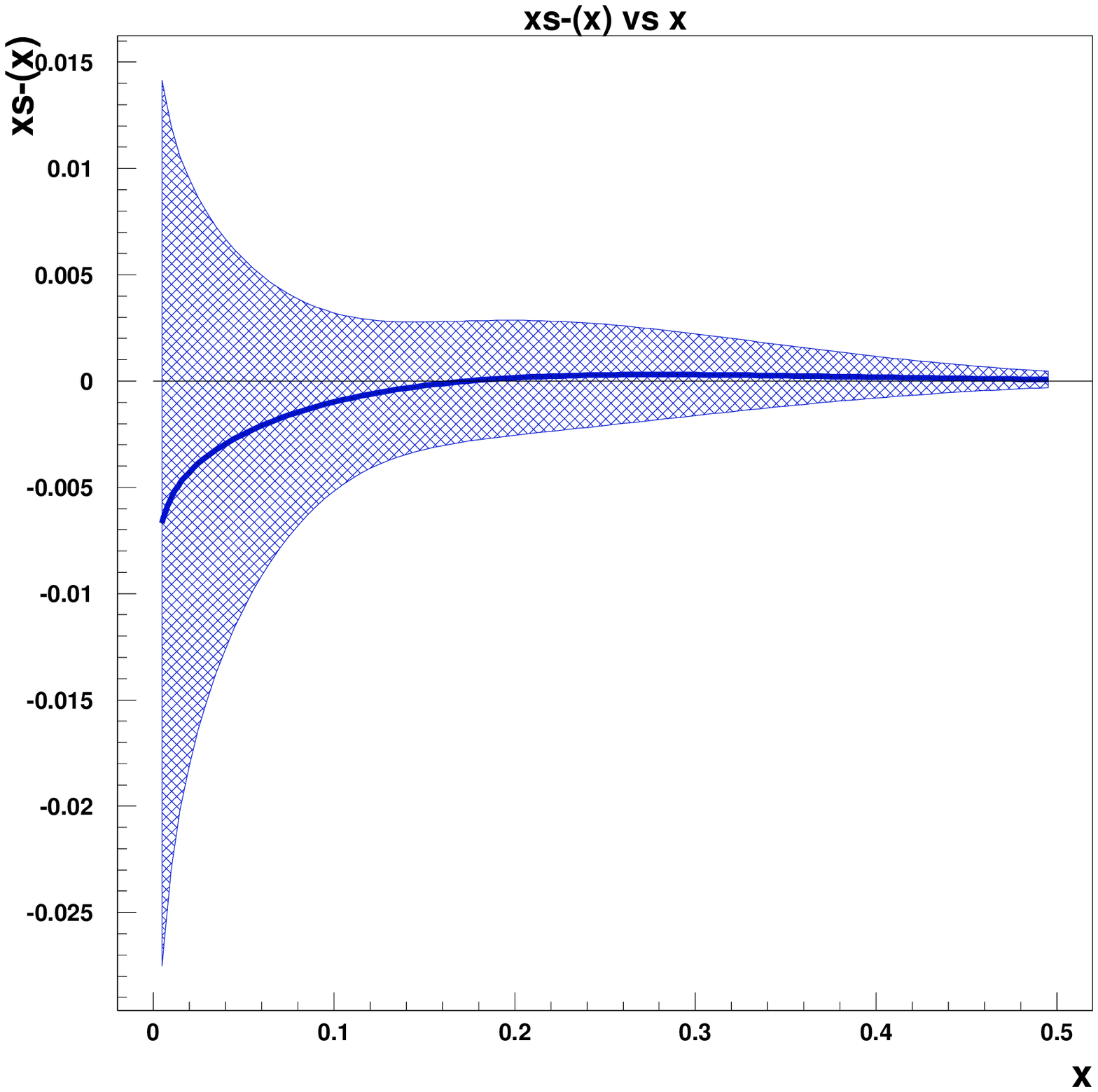}}
\resizebox{3.0in}{2.3in}
   {\includegraphics{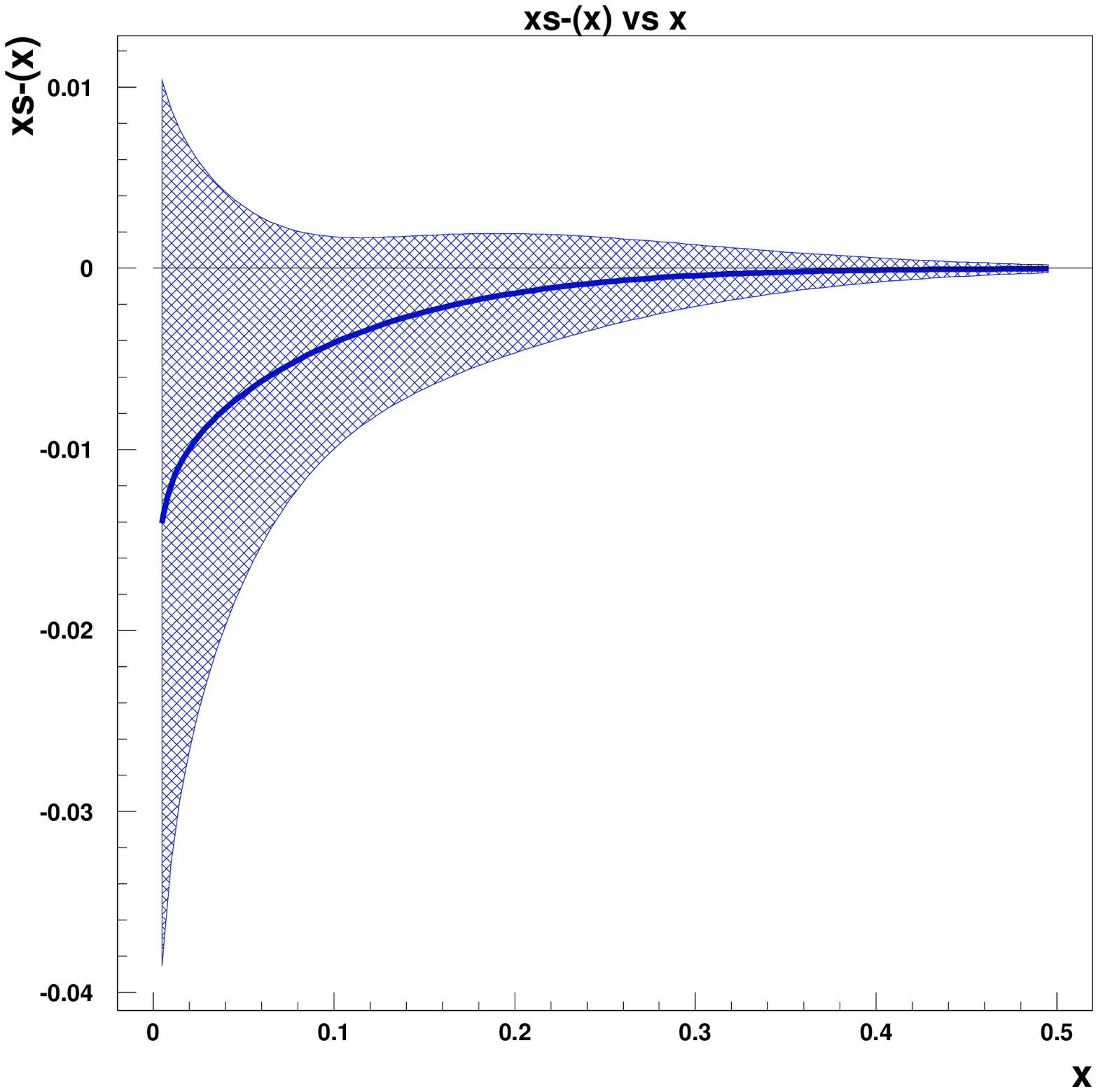}}
\end{center}
\vspace{-0.2in}
\caption{\label{fig:cteq5l} Left: CTEQ5L LO fitted $x(s-\overline{s})$ vs x  Right: NLO CTEQ6M based fit $x(s-\overline{s})$ vs x. }

\end{figure}

QCD evolution is approximated here by modifying the shape of already evolved $\overline{u}$ and $\overline{d}$ distributions through $\kappa$, $\alpha$, $\overline{\kappa}$, $\overline{\alpha}$.   Results from a LO fit based on the CTEQ5L\cite{Lai:1999wy} PDF set are shown in the left half of figure \ref{fig:cteq5l}.  This and all subsequent plots are shown for a typical $Q^2$ of 10 GeV$^2$.  The asymmetry $S^- = \int x s^- dx = \int x[s(x)-\overline{s}(x)] dx$ was found to be $-0.0003 \pm 0.0011$, tending slightly negative, but consistent with zero.  This differs from the published LO result\cite{Rapid} of $-0.0027\pm0.0013$ which was obtained by a LO fit to combined NuTeV and CCFR data with a different set of PDFs.  In this work we are only fitting the NuTeV tables with CTEQ PDFs.

With the LO case as a baseline, an NLO cross section\cite{GOTT,Kretz} fit using CTEQ6M\cite{Pumplin:2002vw} nonstrange PDFs was then performed with the same parameterization as LO.  $S^-$ in this case was $-0.0011\pm0.0014$, again consistent with zero but tending negative at low x.  The asymmetry is plotted on the right half of figure \ref{fig:cteq5l}.  This method technically violates the Alterelli Parisi equations however, by modifying pre-evolved PDFs by an arbitrary function of x.  Another fit was performed with the same functional form for the strange and antistrange seas, this time defining $s(x)$ and $\overline{s}(x)$ at a $Q_0$ of 1.3 GeV.  The PDFs were then properly evolved in $Q^2$ with a version of the EVLCTEQ evolution code that allowed the strange and antistrange seas to be different\cite{TUNG,Olness:2003wz}.  The results obtained were very similar, with an asymmetry of $-0.0013\pm0.0013$.  


Evolving the strange and antistrange seas properly brings the analysis closer to full QCD compliance, however the above scheme still violates the flavor sum rule by not requiring $\int s^-$ to be zero, and violates the momentum sum rule by not normalizing the strange sea relative to the nonstrange PDFs.  Following the example of Olness et al.\cite{Olness:2003wz} a fit was also performed using the following parameterization:

\vspace{-0.2in}
\begin{eqnarray}
  s^+(x,Q_0) & = & \kappa^+ (1-x)^{\alpha^+} x^{\gamma^+} \left[\overline{u}(x,Q_0) + \overline{d}(x,Q_0)\right] \\
  s^-(x,Q_0) & = &  s^+(x) \tanh \left[\kappa^- (1-x)^{\alpha^-} x^{\gamma^-} \left(1-\frac{x}{x_0}\right)\right]\\
  s &= &\frac{s^+ + s^-}{2} ~~~~~ \overline{s} ~~=~~ \frac{s^+ - s^-}{2}
\end{eqnarray}

\begin{figure}[h!]
\begin{center}
\resizebox{3.0in}{2.3in}
   {\includegraphics{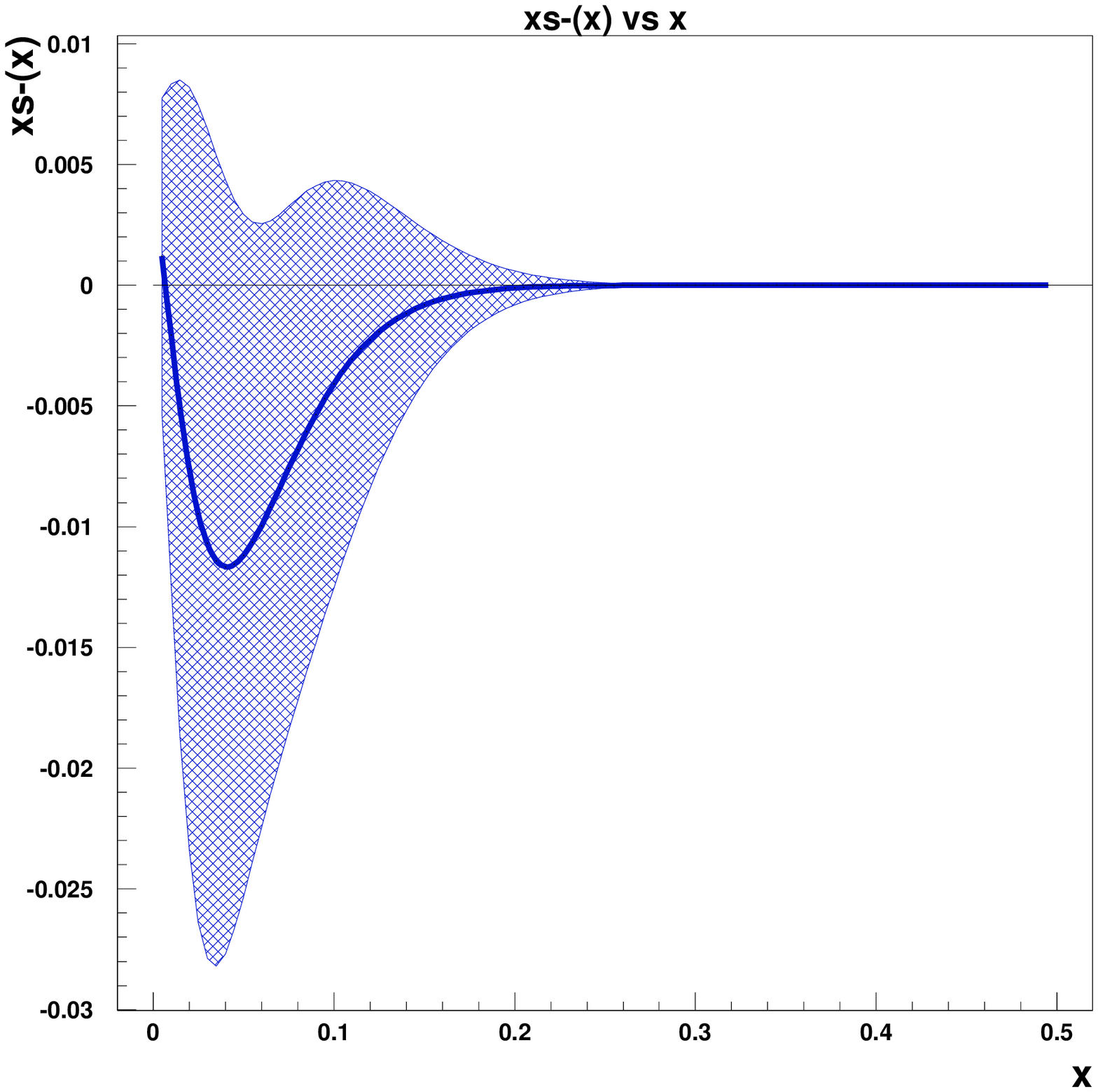}}
\resizebox{3.0in}{2.3in}
   {\includegraphics{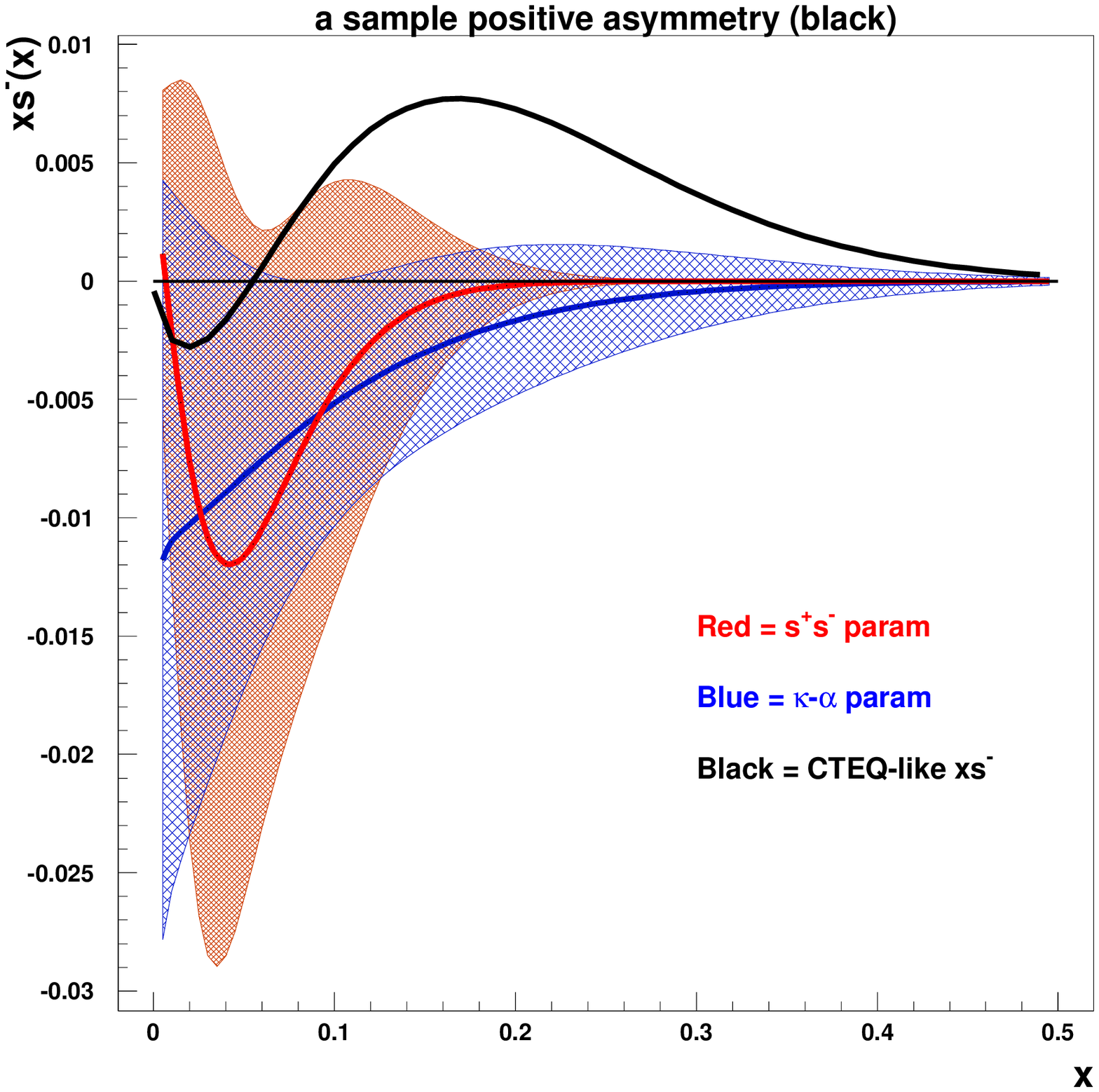}}
\end{center}
\vspace{-0.2in}
\caption{\label{fig:cteqpara} Left: $xs^-(x)$ vs x. from fit with CTEQ $s^+$,$s^-$ scheme.  Right:  Comparing asymmetry curves from fits with $\kappa-\alpha$ parameterization (blue), CTEQ $s^+$,$s^-$ parameterization, and a sample positive asymmetry }
\vspace{-0.0in}
\end{figure}

In this case, the flavor sum rule is enforced by choosing a crossing point $x_0$ such that $\int s^-$ is zero.  Enforcing the momentum sum rule constraint was found not to affect the asymmetry.  Results from fitting with this parameterization are shown in the left half of figure \ref{fig:cteqpara}.  The data prefers satisfying the flavor sum rule by forcing $s^-$ to spike positive below an $x_0$ of 0.009, lower in x than any of the data table points.  The asymmetry follows the same trend as the previous parameterization with an $S^-$ of $-0.0009 \pm 0.0014$.  

To examine whether a positive asymmetry is compatible with the NuTeV dimuon data, an $s^-$ shape similar to that found by the CTEQ collaboration\cite{Olness:2003wz} was held fixed while $s^+$ was allowed to find the best minimum.  The black curve shown in the right half of figure \ref{fig:cteqpara} is the result.  This has a positive $S^-$ of +.0016, less than one quarter of what is needed to resolve the NuTeV $\sin^2\theta_W$ discrepancy.  The best $\chi^2$ of this fit was high, 55/40 NDF.  When $s^-$ was then freed, the asymmetry quickly sought out the shape found on the left of figure \ref{fig:cteqpara} with a $\chi^2$ of 37/37 NDF.

\section{Conclusion}

Several new fit results for the strange asymmetry using the NuTeV forward dimuon cross section table data have been presented.  In general all the fits, LO or NLO, constraining sum rules or not, tended towards a negative strange asymmetry at low x that was still consistent with zero.  Enforcing the flavor sum rule by constraining $\int s^-(x) dx$ to be zero forced $s^-(x)$ to spike positive below where there was data to constrain it.  When the asymmetry was constrained to be positive, the $\chi^2$ increased significantly, making even a small positive asymmetry difficult to accomodate with our data.

These results are still in the process of finalization, and it should be noted that the CTEQ collaboration has reached different conclusions from us in their preliminary global fits including the NuTeV tables.  It is agreed however, that the NuTeV $\sin^2 \theta_W$ discrepancy is unlikely to be the product of an asymmetry in the strange and antistrange PDFs.  NuTeV and CTEQ are continuing an active collaboration to try to resolve any remaining issues and to fully exploit the rich physics possibilities offered by the NuTeV dimuon data.

\end{document}